\documentclass[11pt]{article}

\usepackage{deauthor}
\usepackage{enumitem}
\usepackage{algorithmic}
\usepackage{algorithm}
\usepackage{xcolor}
\usepackage{xspace}
\usepackage{graphicx}
\usepackage{footnote}
\usepackage{hyperref}
\usepackage{threeparttable}
\usepackage{multirow}
\usepackage{subfig}
\usepackage{enumitem}
\usepackage{comment}

\usepackage{times}
\usepackage{wrapfig}
\usepackage{subfig}
\usepackage{booktabs}
\usepackage{url}
\usepackage{amssymb}

\begin{document}

\title{Dimensionality-Reduction Techniques for Approximate Nearest Neighbor Search: A Survey and Evaluation}

\author{Zeyu Wang$^*$, Haoran Xiong\thanks{Co-first authors}, Qitong Wang\footnote{This work was done while Qitong Wang was with Universit{\'e} Paris Cit{\'e}.}, Zhenying He, Peng Wang\thanks{Corresponding author}, Themis Palpanas$^\S$, Wei Wang\\
Shanghai Key Laboratory of Data Science, School of Computer Science, Fudan University\\
  \{wangzeyu17, hrxiong20, zhenying, pengwang5, weiwang1\}@fudan.edu.cn \\
  $^\dagger$Harvard University, qitong@seas.harvard.edu \\
    $^\S$Universit{\'e} Paris Cit{\'e}, themis@mi.parisdescartes.fr}


\setcounter{section}{0}
\setcounter{figure}{0}
\setcounter{table}{0}

\maketitle

\renewcommand\thesection{\arabic{section}}

\begin{abstract}
Approximate Nearest Neighbor Search (ANNS) on high-dimensional vectors has become a fundamental and essential component in various machine learning tasks.
Recently, with the rapid development of deep learning models and the applications of Large Language Models (LLMs), the dimensionality of the vectors keeps growing in order to accommodate a richer semantic representation.
This poses a major challenge to the ANNS solutions since distance calculation cost in ANNS grows linearly with the dimensionality of vectors.
To overcome this challenge, dimensionality-reduction techniques can be leveraged to accelerate the distance calculation in the search process.
In this paper, we investigate six dimensionality-reduction techniques that have the potential to improve ANNS solutions, including classical algorithms such as PCA and vector quantization, as well as algorithms based on deep learning approaches.
We further describe two frameworks to apply these techniques in the ANNS workflow, and theoretically analyze the time and space costs, as well as the beneficial threshold for the pruning ratio of these techniques.
The surveyed techniques are evaluated on six public datasets.
The analysis of the results reveals the characteristics of the different families of techniques and provides insights into the promising future research directions.


\end{abstract}

\section{Introduction}
\label{sec:introduction}

\paragraph{ANN Search}
Approximate Nearest Neighbor Search (ANNS) is a crucial component for numerous applications in various fields~\cite{DBLP:conf/wims/EchihabiZP20}, such as image recognition~\cite{Jang_2020_CVPR}, pose estimation~\cite{1641018}, and recommendation systems~\cite{10.1145/1242572.1242610}, particularly in high-dimensional spaces. Recent studies have shown that deep neural networks, including large language models, can be augmented by retrieval to enhance accuracy~\cite{li2022survey,NEURIPS2020_6b493230} and decrease the magnitude of parameters~\cite{pmlr-v119-guu20a}, further emphasizing the significance of ANNS in modern AI applications. Objects, such as images, documents, and videos, can be transformed into dense vectors in the embedding space. 
ANNS aims to find top-$k$ most similar objects in the embedding space $\mathbb{R}^D$, given a query vector $q\in \mathbb{R}^D$. Compared to the prohibitively high cost of exact search, ANNS is more appealing due to its ability to retrieve high-quality approximate neighbors with a faster response time~\cite{hydra1,hydra2}.
To support efficient ANNS, vector indexes are proposed as special data structures to capture the similarity relation between vectors before querying.
With vector indexes, most of the data that are irrelevant to the query can be quickly pruned when querying, leading to search efficiency.

During ANNS, the time for distance calculation is a major bottleneck.
Taking the popular HNSW~\cite{hnsw} vector index as an example, the distance calculations account for 60\%$\sim$90\% of the total query processing time.
This is also the case for other types of indexes.
The time complexity of common distance metrics, like L2 and inner product, is $\mathcal{O}(D)$ where $D$ is the dimensionality of the vectors.
This indicates that the dimensionality of the vectors is a key factor for the efficiency of ANNS vector indexes.
In public datasets, the vector dimensionality ranges from hundreds to thousands.
With the rapid evolution of pre-trained language models, the dimensionality of embedding vectors also grows: dimensionalities of a few thousands are now commonly used in order to better capture the data semantics~\cite{dubey2024llama, jiang2023mistral, almazrouei2023falcon}.
Some of the latest embedding models, such as Alibaba's Qwen2~\cite{qwen2} and Saleforce's SFR~\cite{SFR}, produce 3584- and 4096-dimensional vectors, respectively.
This presents a significant challenge for the ANNS algorithms.

\paragraph{Dimensionality-Reduction}
To overcome this challenge, dimensionality-reduction techniques can be leveraged to reduce the distance calculation cost.
Specifically, the original high-dimensional vectors can be summarized using lower-dimensional representations, and the distance between these representations can be used to approximate the actual vector distances.
The idea behind this is that the accuracy loss when computing the distance will not necessarily decrease the final search accuracy of ANNS.
In fact, in the top-$k$ nearest neighbor search problem, only the first $k$ vectors require exact distance, while for the other vectors, we only need to confirm they are farther from the query than the top-$k$.
That is, an estimated distance is already sufficient for most distance calculations in the ANNS problem.
Given that the distance estimation achieved by low-dimension representations is usually much more efficient than the full calculation, the performance of current ANNS solutions can be significantly improved.
Moreover, since distance calculation is a basic operation for all ANNS algorithms, this approach can benefit all existing algorithms and is orthogonal to specific index structures.

Nevertheless, the estimated distance provided by current dimensionality-reduction techniques cannot be used to safely prune the non-$k$NN vectors.
That is, the distance of some $k$NN vectors might be over-estimated (and these vectors be skipped), leading to a degraded search accuracy.
Recent methods, like ADSampling~\cite{ads}, leverage random projection as the dimensionality-reduction technique to reduce this estimation error, but with a higher estimation cost.
On the other hand, many dimensionality-reduction techniques are yet unexplored for the ANNS problem.
These techniques, such as Principle Component Analyses (PCA)~\cite{pca} and Discrete Wavelet Transformation (DWT)~\cite{dwt}, are widely studied in the community with solid theoretical foundations, and thus, have the potential to play a positive role in the ANNS problem.

\paragraph{Contributions}
In this paper, we survey six dimensionality-reduction techniques and two frameworks that apply these techniques to the ANNS problem.
The dimensionality-reduction techniques include classical techniques like PCA, machine learning techniques like Product Quantization (PQ)~\cite{pq}, and the deep neural network SEANet~\cite{seanet}.
These techniques show advantages in different scenarios in our experiments.
We analyze these techniques in theory when serving the ANNS problem.
Besides, we study two frameworks to apply these techniques to the ANNS problem, named \emph{in-place transformation} and \emph{out-of-place acceleration}.
The former requires pre-processing for the dataset before building the index and adopts an adaptive query algorithm to reduce the cost of distance calculations.
The latter constructs an auxiliary data structure when building the index and leverages it to accelerate distance calculations.
Furthermore, we implement a pluggable library, named Fudist, to incorporate the dimensionality-reduction techniques as an efficient distance calculator.
With Fudist, we evaluate the techniques on 16 real million-scale datasets of different distributions.
Based on these empirical results, we list open problems on different technical directions in this research area.

All source codes, datasets, and hyper-parameter settings used in our benchmark are available online~\cite{code}.
This ensures the reproducibility of all the experimental results presented in this work.
We hope that Fudist will become a standard library for ANNS research that is orthogonal to the index type and search algorithms, and thus, will help improve the comparison of results from different papers.

The rest of this paper is organized as follows.
We first review the related works in Section~\ref{sec:related}.
In Section~\ref{sec:framework}, we present the two frameworks for applying dimensionality-reduction techniques to the ANNS problem, and in Section~\ref{sec:design}, we describe the surveyed dimensionality-reduction techniques.
The evaluation results are shown in Section~\ref{sec:experiments}, and Section~\ref{sec:conclusion} concludes this paper with a list of open problems.


\section{Related Work and Background}
\label{sec:related}
\paragraph{ANN Indexes}
State-of-the-art ANN indexes~\cite{tkde-benchmark} can be categorized into four classes, proximity-graph-based~\cite{graph-benchmark,wang2023graph}, PQ-based\footnote{\emph
{PQ} stands for \emph
{Product Quantization}.}~\cite{pq,opq}, Locality Sensitive Hashing (LSH)-based~\cite{lsh-apg,pm-lsh} and tree-based~\cite{hd-index,dumpy,dumpyos,odyssey} (though, some hybrid techniques have started to emerge, combining trees with LSH~\cite{DBLP:journals/pvldb/WeiPLP24}, LSH with proximity-graphs~\cite{DBLP:journals/pvldb/ZhaoTHZZ23}, and proximity-graphs with trees~\cite{elpis}; some more recent techniques do not fall in any of these four classes~\cite{suco}).
Among the solutionss in these four classes, graph-based indexes~\cite{knngraphsurvey} achieve the best query performance for $ng$-approximate (i.e., approximate search with no guarantees~\cite{hydra2}) in-memory search, and thus, attracts much interest from the academic and industrial communities.
In this paper, we focus on the popular HNSW~\cite{hnsw} graph-based index, which is a widely adopted index for  ANNS~\cite{faiss,milvus}, and we implement our dimensionality-reduction techniques on HNSW to verify their effectiveness.

\paragraph{Dimensionality-Reduction Techniques}
Dimensionality reduction is an important research problem with several solutions proposed in the literature.
Classical methods include PCA, DWT, MDS~\cite{mds}, Isomap~\cite{isomap}, and MR~\cite{adanns}.
These methods leverage linear or nonlinear transformations to obtain low-dimensional representations.
We select PCA and DWT as representatives since they can be trained efficiently.
Random projection, designed based on Johnson-Lindenstrauss lemma~\cite{larsen2014johnson}, is also widely adopted for dimensionality reduction.
In practice, random projection is usually implemented as the inner products of a vector and a group of random vectors, i.e., Locality Sensitive Hashing (LSH). 
In this paper, we select PM-LSH~\cite{pm-lsh} and ADSampling~\cite{ads} as representatives of this class.

Some machine learning methods can train a codebook to encode vectors as short codewords and then estimate the actual distance with an efficient asymmetric distance calculation~\cite{pq}.
Represented by Product Quantization (PQ)~\cite{pq}, these methods first segment the vector to obtain several subspaces and then cluster sub-vectors on these subspaces.
We select OPQ~\cite{opq} as an optimized version of PQ for evaluation.

Deep neural networks can also train low-dimensional vectors with the loss of distance deviation.
Since no existing models are trained for reducing dimensions in the ANNS problem to the best of our knowledge, we adapt SEANet~\cite{seanet} in ourexperiments to show the potential of this class of methods.



Some other techniques are proposed to reduce vector dimensions for data visualization, like $t$-SNE~\cite{t-sne}, LargeVis~\cite{largevis}, and h-NNE~\cite{sarfraz2022hierarchical}.
A recent survey~\cite{vis-survey} summarizes and evaluates recent progress for these techniques.
However, data visualization usually requires a two- or three-dimensional representation, leading to a significant information loss, which makes these techniques impractical for the ANNS problem.
Space-filling curves, like the Z-order curve and Hilbert-order curve, can also reduce dimensionality by ordering vectors.
Yet, they suffer from the same problem as visualization methods.
Dimensionality-reduction techniques also serve for the training and inference of deep neural networks to reduce memory consumption~\cite{optembed}.
In this case, the target of the reduced representation is to reserve the model accuracy instead of the distance loss~\cite{zhang2023experimental}, which is out of the scope of this paper.

\subsection{Background on Graph-Based Indexes}

We now describe the greedy search algorithm for graph-based indexes, which we will use later on. 

\label{sec:graph-search}

\begin{algorithm}[tb]
\caption{Greedy search (graph $G$, query $q$, entry point $ep$, parameter $ef$)} 
\label{alg:query}
{
\begin{algorithmic}[1]
\STATE $pq$ = a priority queue with unlimited capacity, initialized with $ep$
\STATE $H$  = a max-heap with capacity $ef$ 
\WHILE{$pq$ is not empty}
    \STATE $d_{v_c}, v_c$ = pop an element from $pq$
    \STATE $d_{v_{top}}, v_{top}$ = the heap top of $H$
    \IF{$d_{v_c} > d_{v_{top}}$}
        \STATE break
    \ENDIF
    \FOR{each neighbor $v$ of $v_c$ which has not been accessed}
        \STATE $d_v$ = $dist(v, q)$
        \IF{$d_v < d_{v_{top}}$}
            \STATE Insert $(d_v, v)$ into $pq$ and $H$
        \ENDIF
        \STATE mark $v$ as accessed
    \ENDFOR
    \STATE resize $H$ to be $ef$
\ENDWHILE
\RETURN $k$ smallest elements in $H$
\end{algorithmic}
} 
\end{algorithm}

Graph indexes use a directed graph $G(V,E)$ to index vectors, where each vector is represented as a vertex in $V$, and the edges in $E$ connect vectors based on some kind of proximity.
Graph indexes commonly use greedy search to retrieve $k$NN.
As shown in Algorithm~\ref{alg:query}, the search starts from an entry point $ep$, which is often selected randomly, and then computes the distance between the neighbors of $ep$ to the query $q$.
The accessed points are stored in a priority queue $pq$.
In the next step, the algorithm selects the closest point to $q$ from $pq$ as the next stop to visit and repeats the above process.
Note that not all accessed points can enter $pq$: the algorithm maintains a size-bounded heap $H$ for best-so-far answers, and only the points that are closer than some point in $H$ are qualified to enter $pq$.
Finally, the algorithm terminates when all the points in $pq$ are farther than the points in $H$ to $q$.
The algorithm is greedy because only points that are relatively close to the query can be accessed.
A way to escape such ``local optimuma'' is to increase the capacity of $H$, i.e., $ef$, which is the knob to tune the efficiency-accuracy trade-off in graph indexes.

In this paper, we focus on optimizing the seemingly simple distance calculation step (line 10 in Algorithm~\ref{alg:query}), which is the bottleneck of the query algorithm, as discussed in Section~\ref{sec:introduction}.

\section{Two Frameworks for Making Use of Dimensionality-Reduction} 
\label{sec:framework}
In this section, we introduce two frameworks to leverage different dimensionality-reduction techniques to benefit the search process: \emph{in-place transformation} described in Section~\ref{sec:in-place}, and \emph{out-of-place acceleration} described in Section~\ref{sec:out-of-place}.

\begin{figure}[hbt]
\centering
\subfloat[In-place transformation]{
\centering
\label{fig:in-place} 
\includegraphics[width=.75\linewidth]{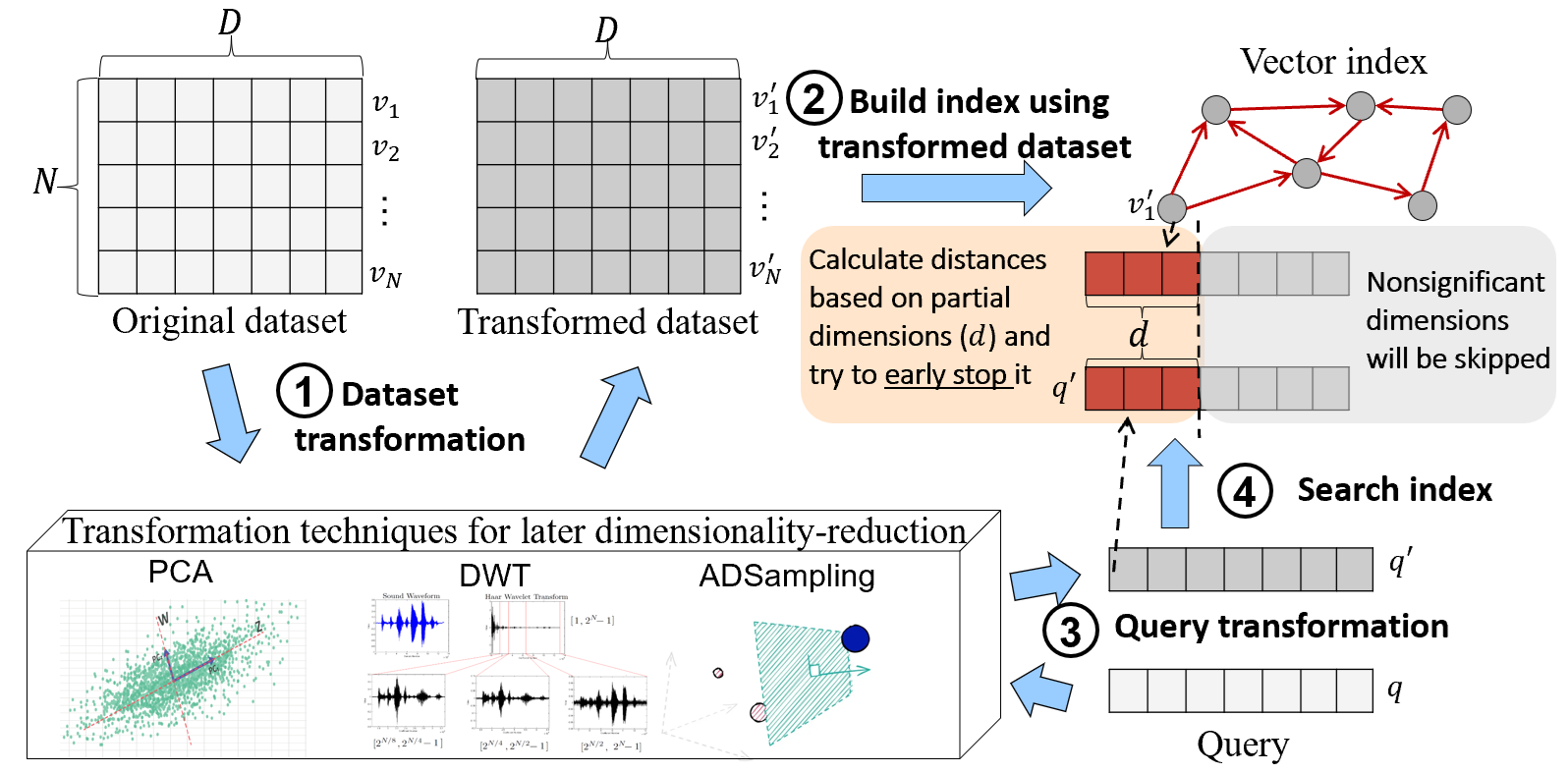}
}

\subfloat[Out-of-place acceleration]{
\label{fig:out-of-place} 
\includegraphics[width=.75\linewidth]{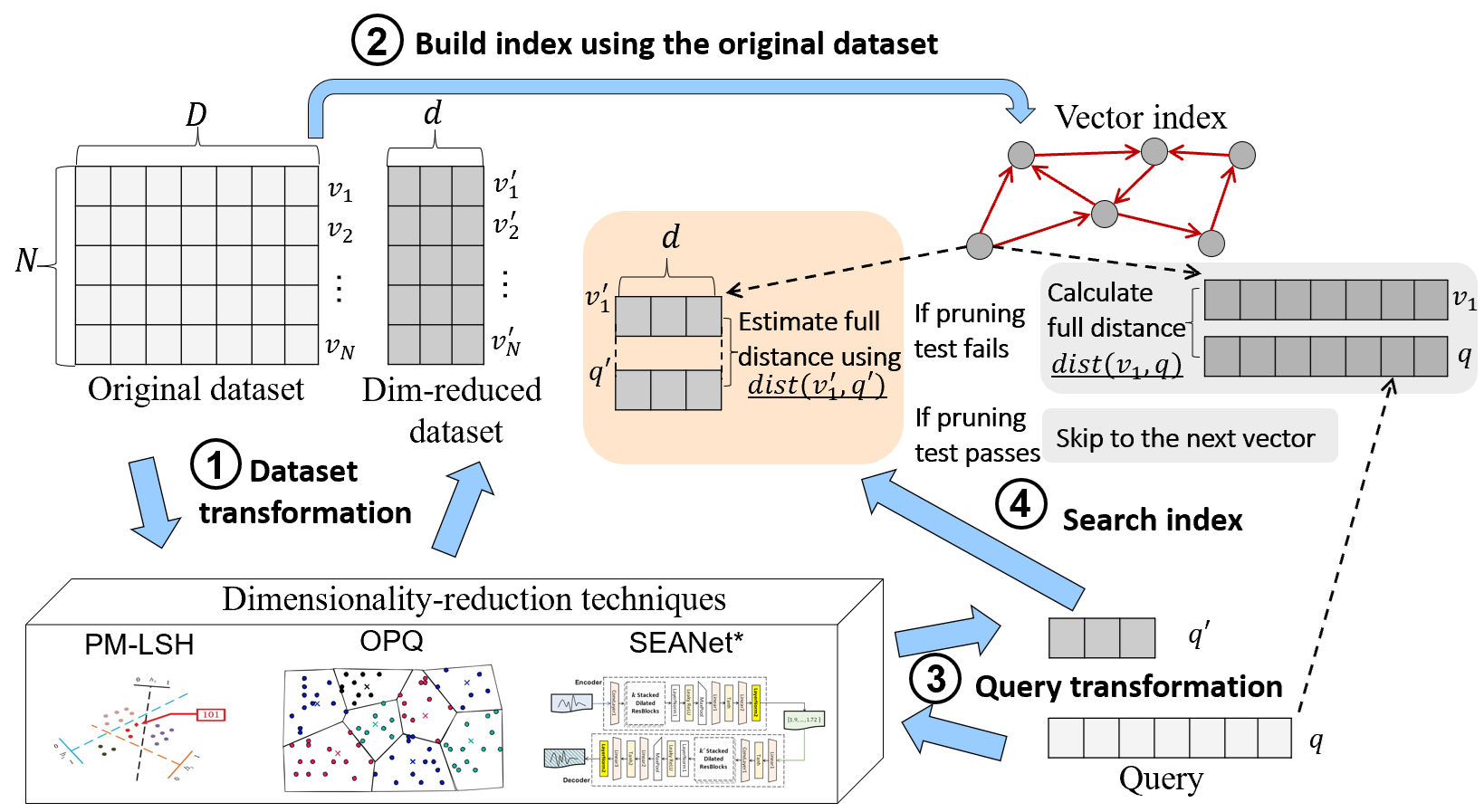}
}
\caption{Illustration of two frameworks to apply dimensionality-reduction techniques 
}
\label{fig:illus} 
\end{figure}

\subsection{In-Place Transformation}
\label{sec:in-place}
The in-place transformation framework was introduced in~\cite{ads}, for a specific dimensionality-reduction technique, ADSampling.
We generalize the framework for all distance-preserved dimensionality-reduction techniques, as shown in Figure~\ref{fig:in-place}.
In this sense, any technique that preserves the distances between vectors after the transformation, can be applied to this framework to benefit the ANNS problem.
For example, all linear transformations, including PCA, are applicable.

In an in-place transformation framework, the dataset and the query will be transformed in a pre-processing step, before indexing and querying.
Then, the vector index will be built based on the transformed dataset.
Note that since vector indexes are constructed based on the distances between vectors rather than the vectors themselves, the vector index built from the transformed dataset will be the same as the original one.
When querying, the query will be first transformed in the same way.
When calculating the distance, the in-place transformation framework adopts an early termination strategy.
That is, we calculate the distance dimension by dimension in a cumulative way.
Since partial distances are smaller than full distance, once the current partial distance has exceeded the threshold, the calculation stops, and the rest of the dimensions are skipped.
The number of computed dimensions will vary for different vectors (see the incarnadine part in Figure~\ref{fig:in-place}).
In the worst case, all the dimensions will be computed and this cost will equal the normal distance calculation cost (besides the cost of termination check).

The rationale behind this method is that transformation techniques will move significant dimensions into the front part of the transformed vector, leading to a higher probabilistic to early terminate the calculation.
To reduce the cost of the termination check, a parameter $\Delta > 1$ is introduced as the cycle to check termination.
$\Delta$ is usually set to 16 or 32 to align with the SIMD instruction width of the distance calculation.

\subsection{Out-of-Place Acceleration}
\label{sec:out-of-place}
The out-of-place acceleration framework leverages an auxiliary data structure, besides the original index structure, to help skip the full distance calculation.
As shown in Figure~\ref{fig:out-of-place}, the vector index is built directly from the raw dataset, and at the same time, a transformed low-dimensional representation table is generated by some dimensionality-reduction technique.
This table is the auxiliary data structure.
When querying, we generate a low-dimensional representation for the query.
For calculating the distance, we first estimate the distance with the low-level representations: if it satisfies the pruning condition, the corresponding full distance will be skipped.
Otherwise, the exact distance should be fully calculated.

In the out-of-place acceleration framework, the dimensionality-reduction technique is not required to preserve exactly the original distances between vectors, or thelower bounds of the original distances.
This means that some vectors might be skipped incorrectly.
Nevertheless, as discussed in Section~\ref{sec:introduction}, only the top-$k$ vectors are required to compute the exact distance, while for the other vectors, dismissal or an approximate distance does not necessarily influence the final search accuracy.
For example, on graph indexes, there could be several paths from the entry point to the destination $k$NN.
A false dismissal might block one path while other paths are still available.
Considering that estimating the distance with low-dimensional representations is usually very efficient, the estimation loss can be compensated by a large  parameter $ef$, which may lead to an overall high search performance.
We also introduce a parameter $\alpha$ as a multiplier of the estimated distance to control the influence of the estimation loss.

\section{Dimensionality Reduction} \label{sec:design}
In this section, we survey six representative dimensionality reduction techniques from various domains that have the potential to benefit the ANNS problem.
The studied techniques are listed in Table~\ref{tab:category}, including three distance-preserved techniques for the in-place transformation framework (i.e., PCA, DWT, and ADSampling), and three other dimensionality-reduction techniques for the out-of-place acceleration framework (i.e., PM-LSH, SEANet*, and OPQ).
In the same table, we also present the additional indexing and query time complexity, as well as the additional space cost.

\subsection{Dimensionlity-Reduction Techniques}
\textbf{PCA}.
Principal Component Analysis (PCA) is a classical linear transformation that selects a new group of vector bases for the data in high-dimensional vector space.
Specifically, the eigenvectors of the dataset are selected as the new basis, and the dimensions with larger eigenvalues (or higher variances) are placed at the front positions of the new vector coordinates.
Therefore, by evaluating the first few dimensions, we expect to obtain most of the distance, and have a high probability of early stopping the full distance calculation.
When applying PCA to the ANNS problem, a transformation matrix is required to reside in memory for pre-processing queries, leading to $D^2$ floats memory cost and $\mathcal{O}(D^2)$ extra query cost.
When calculating distances, we check the termination condition for every $\Delta$ dimension, and thus the extra distance calculation cost is $\mathcal{O}(d/\Delta)$.

Note that PCA can also be leveraged as a dimensionality-reduction technique in the out-of-place acceleration framework.
In this case, we can only take the front part of the transformed dataset and estimate the overall distance with a fixed dimensionality $d$.
However, according to our experiments, using PCA in the out-of-place framework shows inferior performance to that of the in-place transformation framework.
In the rest of this paper, when discussing PCA, we refer to using it in the in-place transformation framework.

\textbf{DWT}.
Discrete Wavelet Transform (DWT) is a classical time series analysis tool following Parseval’s Theorem~\cite{dwt}.
It decomposes the vector using hierarchical wavelet transformations, where the major waves are placed in the first positions. 
The distances between vectors are preserved in both the time and frequency domains, and thus, DWT can be applied in the in-place transformation framework.
Compared to PCA, DWT is not a linear transformation and does not need a matrix to be stored in memory, avoiding the extra memory cost. 
At the same time, it offers linear indexing and querying time costs.
Consequently, DWT is the most lightweight technique among the surveyed techniques.

\begin{table}[tb]
\caption{List of dimensionality-reduction techniques with time and space complexity. $N_c$ is the number of visited points when querying, $d$ is the (expected) dimension of the lower-dimensional representation, $\Delta$ is the termination check cycle, $X$ is the number of neurons in SEANet*, $m$ and $K_s$ are the number of segments and codewords in each codebook of OPQ.}
\label{tab:category}
\centering

\begin{threeparttable}
\begin{tabular}{cccccc}
\toprule
Category &
  Method &
  \begin{tabular}[c]{@{}c@{}}Accuracy\\ guarantee\end{tabular}  &  \begin{tabular}[c]{@{}c@{}}Extra\\query cost\end{tabular} & \begin{tabular}[c]{@{}c@{}}Extra \\indexing cost\tnote{1}\end{tabular}  & \begin{tabular}[c]{@{}c@{}}Extra \\memory cost\end{tabular} \\ \midrule \midrule
\multirow{3}{*}{In-place} & PCA~\cite{pca}  & exact & $\mathcal{O}(D^2+N_cd/\Delta)$  &    $\mathcal{O}(ND^2)$  & $D^2$  \\
                                & DWT~\cite{dwt}    & exact & $\mathcal{O}(D + N_cd/\Delta)$  &        $\mathcal{O}(ND)$  & $0$\\
                                & ADSampling~\cite{ads}    & probabilistic & $\mathcal{O}(D^2+ N_cd/\Delta)$ &   $\mathcal{O}(ND^2)$  & $D^2$  \\ \midrule
\multirow{3}{*}{Out-of-place}     & PM-LSH~\cite{pm-lsh}    & probabilistic & $\mathcal{O}(Dd + N_cd)$ & $\mathcal{O}(NDd)$  & $Nd+Dd$  \\
                                & SEANet*~\cite{seanet}     &     None    & $\mathcal{O}(X + N_cd)$ &   $\mathcal{O}(nX)$ & $\mathcal{O}(X)$ \\                   & OPQ~\cite{opq}     &    None     & $\mathcal{O}(DK_s + N_cm)$         &  $\mathcal{O}(NDK_s)$  & $Nm+DK_s+D^2$  \\ \bottomrule
\end{tabular}
\begin{tablenotes}
\footnotesize
    {
    \item[1] The training time is omitted since the size of the training set is smaller than the dataset.
    }
\end{tablenotes}
\end{threeparttable}
\end{table}

\textbf{ADSampling}.
ADSampling adopts a random squared matrix as the transformation matrix, where each element is sampled from a standard Gaussian distribution.
As indicated in Johnson-Lindenstrauss lemma~\cite{larsen2014johnson}, the random projection has a shape probabilistic error bound for the deviation of the distance.
In this case, the exact distance can be estimated by partial distances with some (probabilistic) confidence.
The more dimensions are calculated, the higher the confidence is.
Therefore, ADSampling has a probabilistic accuracy guarantee instead of a deterministic one (like PCA and DWT).
ADSampling shares the same time and space complexities as PCA, but is much easier to implement, with no training process like SVD in PCA.

\textbf{PM-LSH}
PM-LSH is also designed based on the Johnson-Lindenstrauss lemma~\cite{larsen2014johnson}.
In contrast to ADSampling, PM-LSH directly reduces the dimensionality using the inner product between the vector and a group of random vectors (i.e., the hash function family).
PM-LSH can encode vectors with ultra-low dimensional representations (e.g. 16) and provides an accuracy guarantee for the distance deviation.
Similar to ADSampling, PM-LSH requires a random transformation matrix in memory when querying, leading to $Dd$ space cost and $\mathcal{O}(Dd)$ query pre-processing cost.
In addition, in the out-of-place acceleration framework, the dimensionality-reduced dataset (i.e., the data after hashing) requires $Nd$ space.
To generate this auxiliary structure, we additionally need a hashing operation for the dataset when building the index, with $\mathcal{O}(NDd)$ time cost.
When calculating a distance, estimating the distance with the low-dimensional representations is necessary, for an extra $\mathcal{O}(d)$ cost.
If the pruning condition is not satisfied, the full distance will be calculated.

\textbf{SEANet*}.
Series Approximation Network (SEANet) is a deep neural network proposed to represent high-frequency data series with deep learning embeddings.
These embeddings are further indexed by the iSAX tree index family~\cite{isax,DBLP:journals/vldb/PengFP21,odyssey,dumpy,dumpyos}.
SEANet adopts an encoder-decoder framework with a loss function comprising both distance deviation and vector reconstruction error.
We adapt SEANet to the ANNS problem, resulting in SEANet*.
Specifically, we remove the decoder part to make SEANet* an encoder-only framework, since vector reconstruction is not necessary in the ANNS problem.
Moreover, we remove the construction error in the loss function to focus on distance preservation.
Training SEANet introduces a significant time cost when building the index, and it also incurs higher space costs than other alternatives, in order to store the network.
Nevertheless, as we will show in the experiments, SEANet* displays a strong potential to improve the query performance significantly.

\textbf{OPQ}.
Product Quantization (PQ) is a popular vector compression technique that is commonly adopted along with the IVF index (i.e., iVF-PQ~\cite{faiss}).
In the context of a graph-based index, PQ helps direct the search in memory, in order to reduce the I/O cost of a disk-based index, DiskANN~\cite{diskann}.
In this paper, we use it to estimate distances efficiently.
During indexing, we train the codebooks of the dataset and obtain the codewords of vectors with the codebooks as the auxiliary data structure.
When querying, we first generate a distance look-up table using the query and the codebooks.
When calculating the distance, the distance can be efficiently estimated by looking up the distance table.
Note that OPQ does not give any accuracy guarantee on the distance estimation.
Nevertheless, due to the efficiency of distance estimation, the accuracy loss can often be compensated, resulting in an overall performance improvement.


\subsection{Beneficial Threshold}
We now study at which point the dimensionality-reduction techniques can improve the query performance of the ANNS algorithm, instead of hurting it.
Specifically, we focus on a key factor, the pruning ratio $\rho$, defined as $\rho=1 - \frac{N_f}{N_c}$, where $N_f$ is the number of full distance calculations and $N_c$ is the number of visited points.

To make our methods more efficient than the raw HNSW, the following inequation should hold:
\begin{equation}
    N_c\cdot (C_e + (1-\rho)\cdot\mathcal{O}(D)) + C_p < N_c'\cdot \mathcal{O}(D),
\end{equation}
where $C_e$ and $C_p$ are the distance estimation cost and query pre-processing cost, respectively (i.e., the second and the first term of the fourth column in Table~\ref{tab:category}), and $N_c'$ is the number of visited points for the raw HNSW to achieve the same query accuracy.
We can derive the \emph{beneficial threshold} of the pruning ratio $\rho$:
\begin{equation}
\label{equ}
    \rho > 1 -  \frac{N_c'}{N_c} + \frac{1}{\mathcal{O}(D)}(\frac{C_p}{N_c} + C_e ). 
\end{equation}

The second term describes the additional search cost of our methods, compared to the raw HNSW, because of the distance estimation deviation.
For distance-preserved methods, this term is equal to 1 in theory.
For others, this term is smaller than 1.
However, in practice, we observe that even for distance-preserved methods, this term is also a bit smaller than 1, due to the transformation and calculation error of float-point numbers.
The third term describes the ratio of the amortized cost of distance estimation to the full calculation for each vector.
Obviously, with a more accurate and efficient dimensionality-reduction technique, the beneficial threshold will be lower.

\section{Experiments}  \label{sec:experiments}

\subsection{Experimental Settings}
\paragraph{Setup}
Experiments were conducted on an Intel(R) Xeon(R) CPU E5-2620 v4 @ 2.10GHz CPU 20MiB L3 cache with 128GB 2400MHz main memory, running Ubuntu Linux 16.04 LTS. All codes are implemented in C++ and compiled in g++ 9.4.0 with -O3 optimization. 
SIMD operations are implemented with AVX instructions. The codes are open-sourced in~\cite{code}.

\paragraph{Datasets} \label{subsec:datasets}
We select six public datasets of different dimensionality and hardness~\cite{steiner}, as listed in Table~\ref{tab:datasets}.
We generate unbiased workloads consisting of queries with different hardness using the $Steiner$-Hardness method~\cite{steiner} for datasets GIST, H\&M, Imagenet, and Deep.
We use the public workloads for Trevi and MNIST (their data distributions make them hard to augment with the $Steiner$-Hardness method). 

\begin{table}[tb]
  \caption{Dataset characteristics}
  \label{tab:datasets}
  \centering
  \begin{tabular}{llllll}
    \toprule
    Datasets  & Dataset size & Query size   & D  & Hardness\\
    \midrule
    Trevi & 100,000 & 200 & 4096 & 5335   \\
    H\&M & 105,100 & 10,000 & 2048 & 3439 \\
    GIST & 1,000,000  & 10,000 & 960 & 12381    \\
    MNIST & 69,000 & 200 & 784 & 1010  \\
    Imagenet & 1,000,000 & 10,000 &150 & 10240    \\
    Deep     & 1,000,000 &  10,000 &96 & 9451  \\
    \bottomrule
  \end{tabular}
\end{table}

\paragraph{Metrics}
We use recall to measure the accuracy of ANNS.
Formally, $Recall\,k@k=\frac{|A\cap G|}{k}$, where $A$ is the returned approximated neighbors sets and $G$ is the ground truth (i.e., real $k$NN).

\paragraph{Hyper-parameters}
$k$ is set to 20, and queries are executed using one thread.
The construction parameters of HNSW, $M$ and $efCons$, are tuned to achieve the best query performance.
The multiplier $\alpha$ is tuned to make each dimensionality-reduction technique in the out-of-place acceleration framework perform the best.
DWT requires the length of the vector to be a power of two, while OPQ requires it to be a multiple of the number of segments.
For these two methods, we add padding zeros to the datasets that do not satisfy the corresponding requirements.
For DWT, we remove the all-zero columns after thetransformation.
For ADSampling, we fix the hyper-parameter $\epsilon_0$ to be 2.1 as recommended, which usually reaches the optimum in our experiments.

\begin{figure}[bt]
\subfloat[Trevi (D=4096)]{
\label{fig:ann-trevi} 
\includegraphics[width=0.3\linewidth]{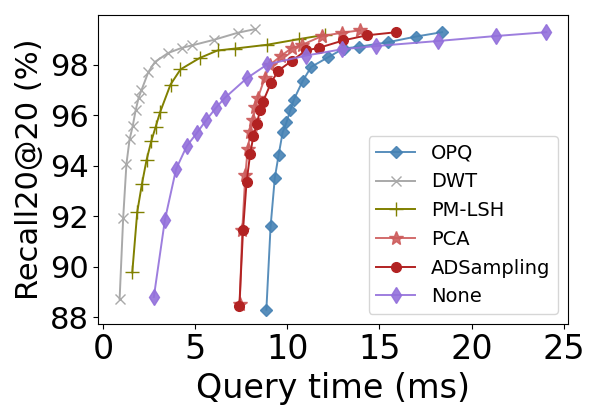}
}
\subfloat[H\&M (D=2048)]{
\label{fig:ann-word2vec} 
\includegraphics[width=0.3\linewidth]{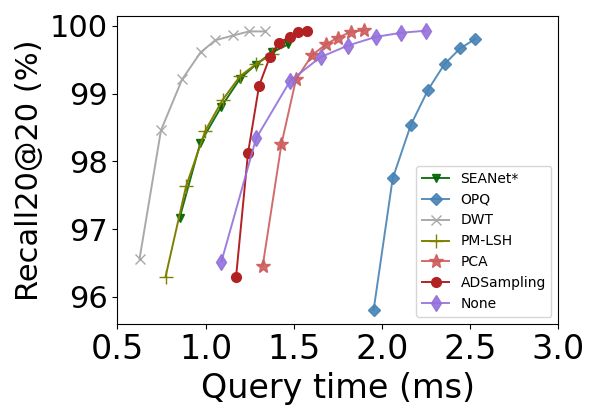}
}
\subfloat[GIST (D=960)]{
\label{fig:ann-gist} 
\includegraphics[width=0.3\linewidth]{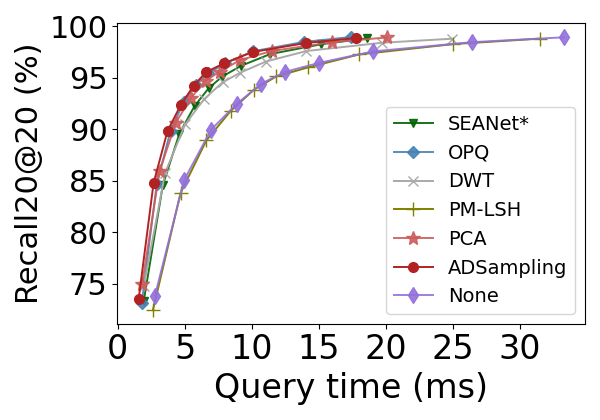}
}

\subfloat[MNIST (D=784)]{
\label{fig:ann-mnist} 
\includegraphics[width=0.3\linewidth]{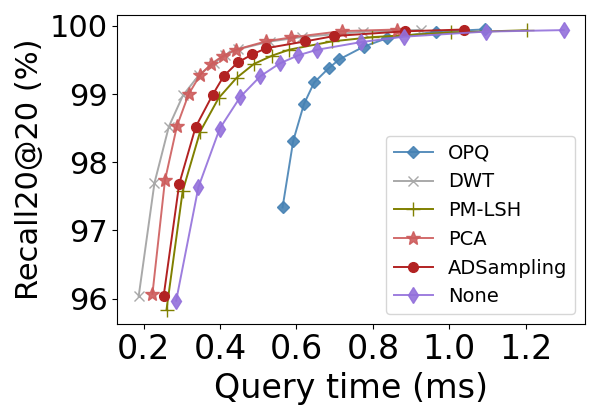}
}
\subfloat[Imagenet (D=150)]{
\label{fig:ann-glove} 
\includegraphics[width=0.3\linewidth]{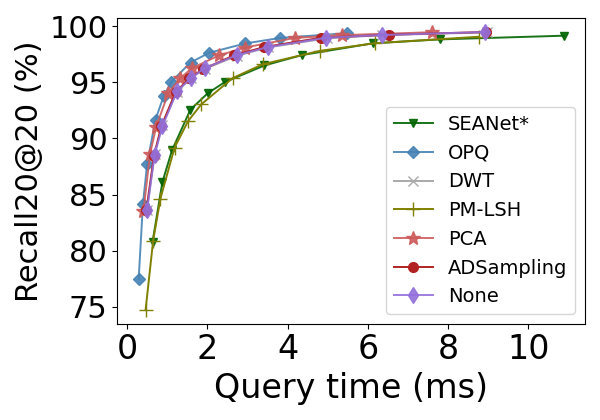}
}
\subfloat[Deep (D=96)]{
\label{fig:ann-deep} 
\includegraphics[width=0.3\linewidth]{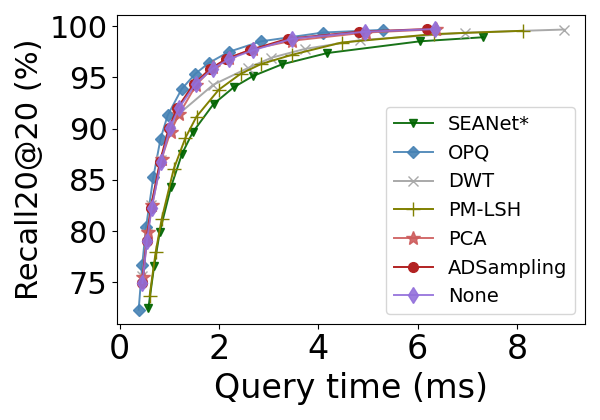}
}
\caption{Query performance with dimensionality-reduction techniques.}
\label{fig:ann} 
\end{figure}

\subsection{Search Performance}
We first disable the utilization of SIMD instructions for distance calculations as in~\cite{ads}, and test the overall search performance of the dimensionality-reduction techniques.
The results are shown in Figure~\ref{fig:ann}.
For these six datasets, the best alternatives improve the performance of raw HNSW by 6.3x, 2.1x, 2.0x, 1.6x, 1.1x, and 1.1x, respectively under 98\% recall, where the best methods are respectively DWT, ADSampling, DWT, and OPQ for the rest three.
The performance improvement significantly increases when the dimensionality of the dataset grows to over 700, where the benefit of skipping one-time full-distance calculation is larger.
The result of SEANet* on Trevi and MNIST is omitted since the model does not coverage on very high-dimensional (Trevi) and sparse (MNIST) datasets. 

On the very high-dimensional dataset Trevi, DWT, and PM-LSH provide significant improvement thanks to their fast preprocessing time, with a complexity of $\mathcal{O}(D)$ and $\mathcal{O}(Dd)$, respectively.
On a simple, but with high-dimensional dataset, preprocessing plays an important role in the overall performance.
Other methods can only provide an improvement on the high-recall range ($>$98\% recall), where more distance calculations can amortize the preprocessing cost.

A similar behavior is observed in the H\&M dataset in Figure~\ref{fig:ann-word2vec}, where DWT and PM-LSH perform the best again.
SEANet* achieves outstanding performance that is close to PM-LSH, as well. since the data distribution of this dataset can be well described by the neural network model.
OPQ cannot improve the original HNSW performance on this dataset due to its long pre-processing time.

On the GIST dataset, ADSampling, SEANet*, PCA, OPQ, and DWT all show about 1.5x performance improvement over the raw HNSW. 
Only PM-LSH loses its edge on this dataset.
The GIST dataset is a good representative of modern vector embedding datasets, which are dense and have high dimensionality with medium hardness.

MNIST is a sparse dataset where most of the values in the vectors are zeros.
In this case, techniques that can effectively summarize the key information like DWT and PCA, show obvious advantages, while OPQ does not perform well since clustering on this sparse dataset tends to be ineffective.
Random projections like ADSampling and PM-LSH are generally inferior to DWT and PCA.

On the rest three low-dimensional datasets, OPQ, PCA, and ADSampling show no more than 20\% improvements over HNSW, while the other three methods show marginally positive (see Figure~\ref{fig:ann}(d) and (e)) or even negative (see Figure~\ref{fig:ann}(f)) influence on HNSW.

Moreover, we observe that not all techniques outperform the raw HNSW; this is true for all datasets we used in our experiments.
This indicates that on some datasets, the pruning ratio of some dimensionality-reduction techniques cannot reach the beneficial threshold. 
Since no single method consistently outperforms all others, selecting the best technique for different datasets is a necessary step. 

\begin{figure}[bt]
\subfloat[Trevi (D=4096)]{
\label{fig:ann-trevi} 
\includegraphics[width=0.3\linewidth]{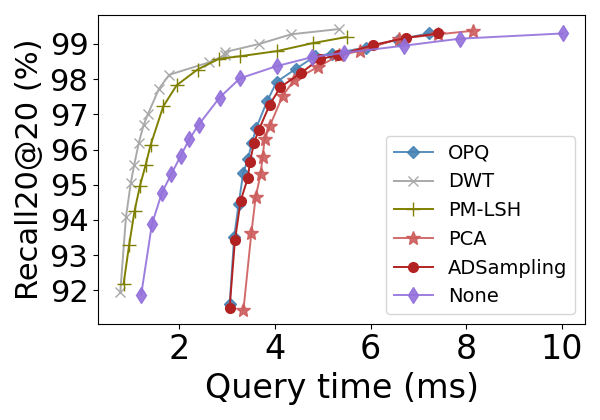}
}
\subfloat[H\&M (D=2048)]{
\label{fig:ann-word2vec} 
\includegraphics[width=0.3\linewidth]{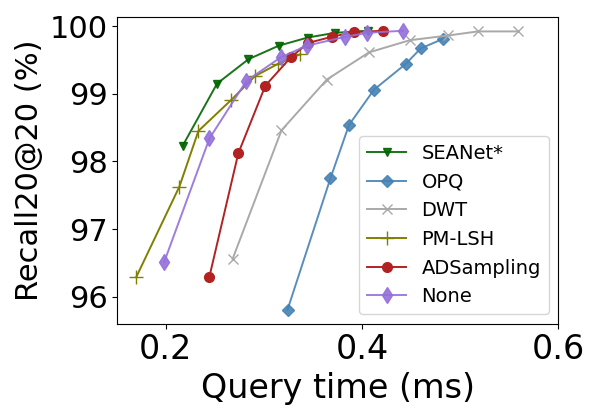}
}
\subfloat[GIST (D=960)]{
\label{fig:ann-gist} 
\includegraphics[width=0.3\linewidth]{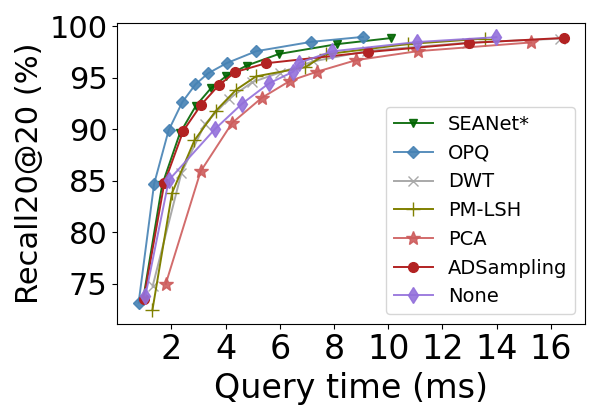}
}

\subfloat[MNIST (D=784)]{
\label{fig:ann-mnist} 
\includegraphics[width=0.3\linewidth]{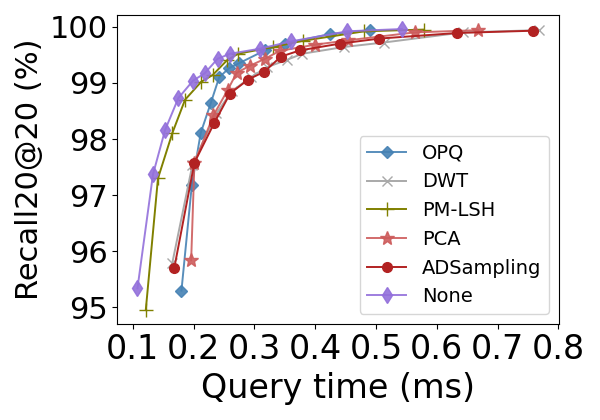}
}
\subfloat[Imagenet (D=150)]{
\label{fig:ann-glove} 
\includegraphics[width=0.3\linewidth]{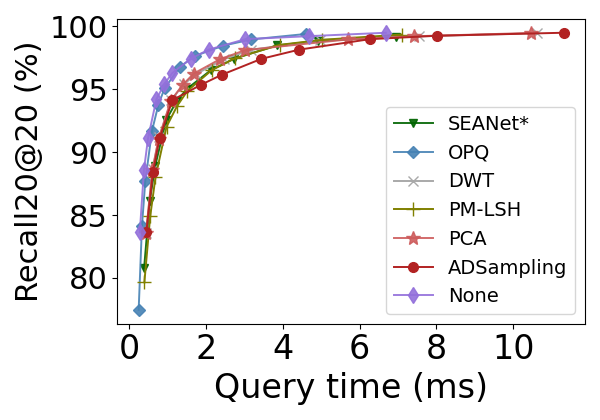}
}
\subfloat[Deep (D=96)]{
\label{fig:ann-deep} 
\includegraphics[width=0.3\linewidth]{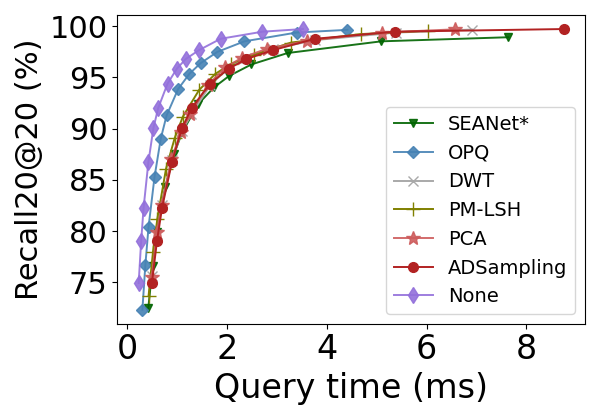}
}
\caption{ANNS performance with SIMD (AVX)}
\label{fig:simd-ann} 
\end{figure}

\paragraph{Search with SIMD instructions.}
Since SIMD instructions are widely adopted in ANNS libraries, such as FAISS and hnswlib, we also test the performance improvement of the surveyed techniques with SIMD (AVX) instructions.
The results are shown in Figure~\ref{fig:simd-ann}.
For these six datasets, the average query speedup of the best alternatives is 2.2x, 1.2x, 1.7x, 0.9x, 0.9x, and 0.8x, compared to the raw HNSW, for 98\% recall. 
The respective best method is DWT, OPQ, PM-LSH, and OPQ for the rest three.

On the Trevi dataset, all methods lead to an improvement, albeit a small one.
OPQ improves more than PCA, since the implementation of the quantization techniques fully leverages the SIMD instructions.
On the H\&M dataset, only SEANet* improves over the original HNSW across the whole recall range.
On the GIST dataset, OPQ isthe best alternative thanks to its high estimation efficiency.
SEANet* comes second after OPQ, followed by ADSampling, PM-LSH, and DWT.
PCA performs worse than the raw HNSW in this case.

We note that when the dimensionality is smaller than 800 (this includes the MNIST, Imagenet, and Deep datasets), the techniques we evaluated cannot provide a significant improvement over the raw HNSW, since the SIMD acceleration of the raw distance calculations reduces the benefit of the additional pruning that these techniques offer.

\subsection{Accuracy Loss}

\begin{figure}[bt]
\subfloat[All DCOs]{
\label{fig:app-ratio} 
\includegraphics[width=0.4\linewidth]{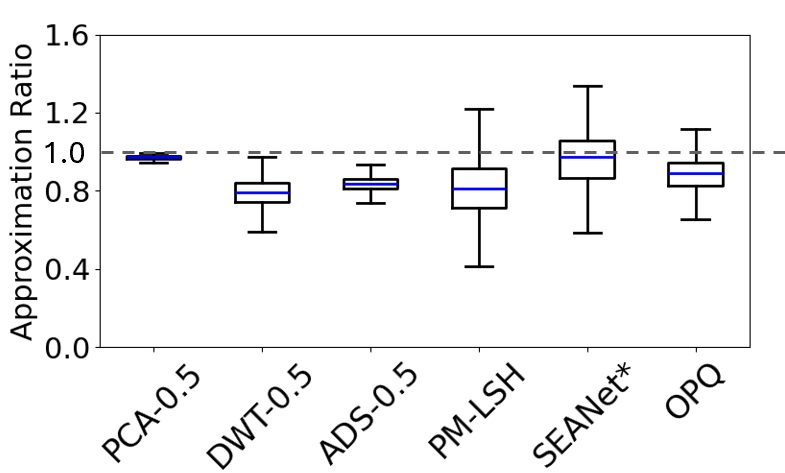}
}
\subfloat[Adaptive DCOs]{
\label{fig:app-ratio-inc} 
\includegraphics[width=0.52\linewidth]{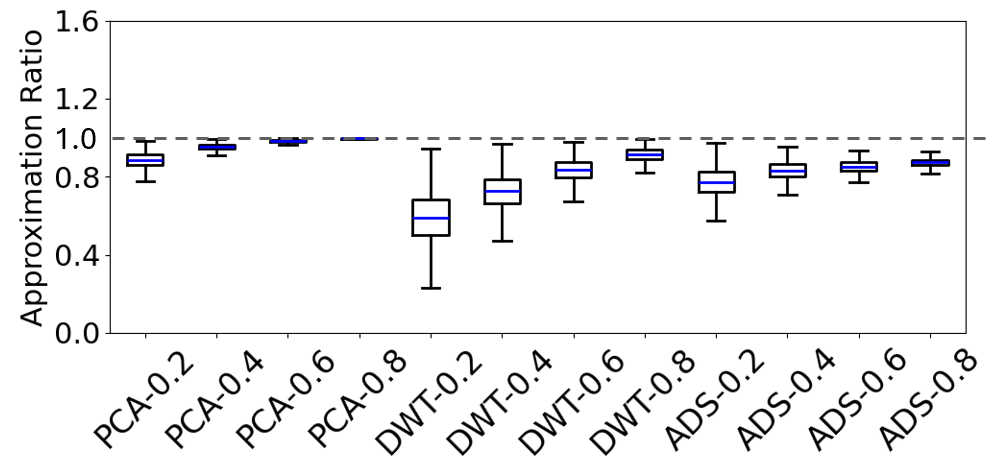}
}

\subfloat[All DCOs]{
\label{fig:app-ratio2} 
\includegraphics[width=0.4\linewidth]{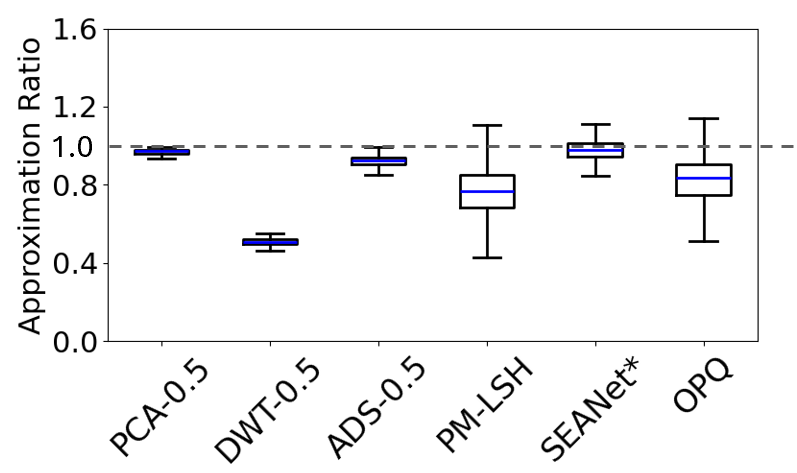}
}
\subfloat[Adaptive DCOs]{
\label{fig:app-ratio-inc2} 
\includegraphics[width=0.52\linewidth]{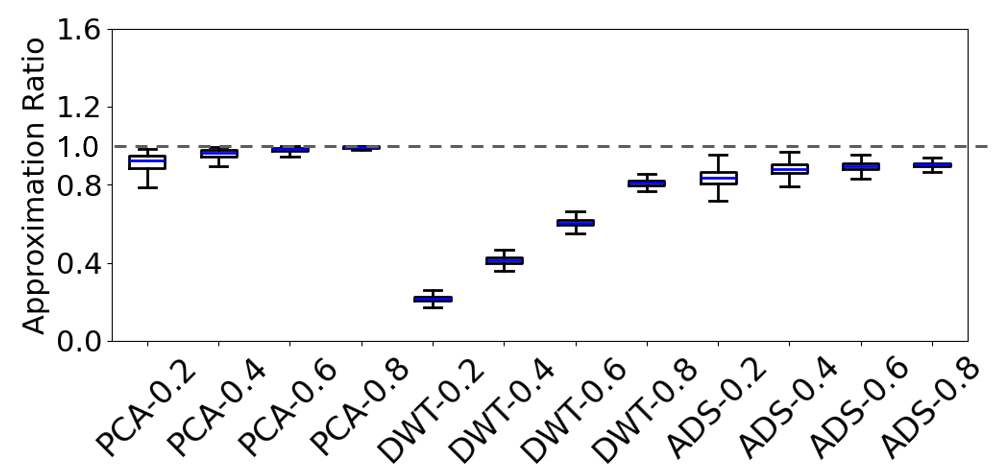}
}
\caption{Approximation ratio on GIST (top) and H\&M (bottom) }
\label{fig:approx} 
\end{figure}

In this section, we study the characteristics of the surveyed dimensionality-reduction techniques by calculating the Approximation Ratio, defined as $AR = \frac{\widehat{dist}(v,q)}{dist(v, q)}$ where $\widehat{dist}$ is the estimated distance.
We sample partial vectors from training and test sets to evaluate the approximation ratios.

Figure~\ref{fig:approx} depicts the distribution of $AR$ on GIST and H\&M datasets, where the dotted line indicates $AR=1.0$, an optimal case of distance-estimation techniques.
We use the notation PCA-0.5 to indicate that weestimate distances using the first 50\% of the dimensions in the transformed dataset.

As distance-preserved techniques, the $AR$ of PCA and DWT is always below 1; PCA is much tighter than DWT, with a smaller variance.
Similarly, although ADSampling does not provide a lower bound, it hardly generates false dismissals thanks to itsreliable probabilistic guarantee.
On the H\&M dataset (refer to  Figure~\ref{fig:app-ratio2}), SEANet* provides much more accurate estimation than PM-LSH and OPQ.
This result indicates the strong potential of deep learning methods to estimate the similarity between very high-dimensional vectors by using  low-dimensional representations.
As shown in Figures~\ref{fig:app-ratio-inc} and ~\ref{fig:app-ratio-inc2}, when applying dimensionality-reduction techniques in the in-place transformation framework, we can expect a more accurate estimation with a smaller variance by checking more dimensions.
For PCA, we can already get a very accurate estimation by calculating 80\% of the dimensions when querying.
Although DWT shows inferior performance than the other techniques, with no more than 50\% of the dimensions, it quickly turns accurate with more dimensions ($>$60\%).

\subsection{Verification of Beneficial Threshold}

\begin{table}[]
\caption{Key metrics evaluation on Deep and GIST on recall = 0.94. $N_c'$ is the number of visited points for the raw HNSW, and $\mathcal{O}(D)$ is the cost of full distance calculation. $N_c$ is the number of visited points for corresponding dimensionality-reduction techniques, $C_e$ and $C_p$ are the distance estimation cost and query pre-processing cost, $\rho$ and $\theta$ are the pruning ratio and the beneficial threshold.}
\label{tab:perform}
\centering
\small
\begin{tabular}{l|ccccc|ccccc}
\toprule
\multirow{2}{*}{Recall@0.94} & \multicolumn{5}{c|}{\textbf{Deep}. $N_c^\prime=6760.3$, $\mathcal{O}(D)\approx 0.2423$ $\mu s$}& \multicolumn{5}{c}{\textbf{GIST}. $N_c^\prime=7929.5$, $\mathcal{O}(D) \approx 1.305$ $\mu s$}\\ \cmidrule{2-11}
                             & \multicolumn{1}{c}{$N_c$} & \multicolumn{1}{c}{$C_e$ ($\mu s$)} & \multicolumn{1}{c}{$C_p$ ($\mu s$)} & \multicolumn{1}{|c|}{$\rho$}        & \multicolumn{1}{c|}{$\theta$} & \multicolumn{1}{c}{$N_c$} & \multicolumn{1}{c}{$C_e$ ($\mu s$)} & \multicolumn{1}{c}{$C_p$ ($\mu s$)} & \multicolumn{1}{|c|}{$\rho$}        & \multicolumn{1}{c}{$\theta$} \\ \midrule
ADSampling                   & 6760.1                 & 0.2354                      & 2.939                       & \multicolumn{1}{|r|}{0.9301} & \textbf{0.9734}                & 8165.3                 & 0.6327                      & 194.4                       & \multicolumn{1}{|r|}{\textbf{0.9383}} & 0.5319               \\
DWT                          & 6813.1                 & 0.2348                      & 1.066                       & \multicolumn{1}{|r|}{0.9423} & \textbf{0.9775}                & 7812.2                 & 0.8935                      & 6.836                       & \multicolumn{1}{|r|}{\textbf{0.9399}} & 0.6703               \\
PCA                          & 6821.7                 & 0.2420                      & 2.956                       & \multicolumn{1}{|r|}{0.9424} & \textbf{1.0093}                & 7773.3                 & 0.7439                      & 195.2                       & \multicolumn{1}{|r|}{\textbf{0.9402}} & 0.5691               \\
PM-LSH                          & 7185.4                 & 0.1300                      & 0.681                       & \multicolumn{1}{|r|}{0.1601} & \textbf{0.5960}                & 8538.8                 & 0.2200                      & 7.713                       & \multicolumn{1}{|r|}{\textbf{0.2938}} & 0.2406               \\
SEANet*                       & 8433.9                 & 0.1230                      & 14.48                       & \multicolumn{1}{|r|}{0.2616} & \textbf{0.7131}                & 7836.9                 & 0.1471                      & 16.59                       & \multicolumn{1}{|r|}{\textbf{0.4828}} & 0.1024               \\
OPQ                          & 7148.8                 & 0.0768                      & 19.22                       & \multicolumn{1}{|r|}{\textbf{0.5703}} & 0.3832                & 8274.5                 & 0.1194                      & 371.9                       & \multicolumn{1}{|r|}{\textbf{0.5941}} & 0.1676               \\ \bottomrule
\end{tabular}
\end{table}


In the final experiment, we calculate each term in Equation~\ref{equ} to verify the effectiveness of the beneficial threshold, and evaluate the practical cost of the asymptotic complexities reported in Table~\ref{tab:category}.
We report the results on the Deep and GIST datasets, in Table~\ref{tab:perform}.
On the Deep dataset, it is only for OPQ that the pruning ratio $\rho$ exceeds the beneficial threshold $\theta$, while on the GIST dataset, all the methods meet the threshold.
This verifies the experimental results shown in Figure~\ref{fig:ann}.
On the Deep dataset, the beneficial thresholds for distance-preserved techniques are very high, and for PCA it is larger than 1, which indicates it is impossible to gain performance improvement.
On the GIST dataset, the increase of the practical cost of $\mathcal{O}(D)$ leads to a major drop of the beneficial threshold $\theta$.
Moreover, according to the value of $1 -  \frac{N_c'}{N_c}$, we observe that the extra search cost introduced by the accuracy loss occupies a small portion of $\theta$.
Except for the very high-dimensional and simple dataset Trevi, the amortized preprocessing cost, $\frac{C_p}{N_c\mathcal{O}(D)}$ is also very small.
In this case, the estimation cost $\frac{C_e}{\mathcal{O}(D)}$ is the most important factor.
We observe that OPQ wins the first place on both datasets w.r.t. the estimation efficiency.
We also note that on the GIST dataset, the $N_c$ value of SEANet* is close to that of the distance-preserved methods, which indicates a strong potential for the estimation effectiveness of deep learning methods.


\section{Conclusions and Future Directions} \label{sec:conclusion}

In this paper, we survey six dimensionality-reduction techniques that have the potential to benefit the query performance of the ANNS algorithm.
We employ two frameworks, in-place transformation, and out-of-place acceleration, to integrate these techniques into the ANNS indexing and querying workflow.
Under these frameworks, we study the theoretical time and space complexity and benchmark their performance with an extensive and fair evaluation.
The results indicate that the best alternatives improve the query performance of the original HNSW by up to 6x, but at the same time, the performance of each technique varies widely across different datasets, and in some cases, we observe no improvement at all.

We observe that at the framework level, the out-of-place acceleration framework offers greater flexibility for use with pre-existing graph indexes, as it can enhance query performance through the addition of an auxiliary data structure. 
In contrast, the in-place transformation framework necessitates the reconstruction of the index, but with lower memory usage compared to the out-of-place framework.

Based on the results of our study, we discuss below promising research directions.

\textbf{1. Quantization techniques with quality guarantees.} Due to the high efficiency of distance estimation, OPQ shows a robust performance improvement in the high-recall range.
However, since the distance estimation does not come with any guarantees, it takes much time to tune the parameters to achieve the optimal trade-off between efficiency improvement and accuracy loss.
In this case, an accuracy guarantee, which can be provided by LSH and the techniques of the in-place transformation framework, will make PQ more feasible and efficient for the ANNS problem~\cite{rabitq,rabitqe}.

\textbf{2. Deep neural networks show a strong potential for efficient low-dimensional representations.}
Under the same dimensionality, deep learning methods show the best estimation accuracy over all methods on some datasets. 
There is still a large design space for deep learning methods w.r.t. the model framework, loss function, sampling, and training strategy, in this scenario.
Moreover, finding the optimal hyper-parameters, such as the dimensionality of the representations, is a challenging open problem.

\textbf{3. Adaptive dimensionality-reduction techniques selection.} 
There is no single technique outperforming all the others according to our evaluation.
This indicates that an effective method selection approach is necessary in practice.
One of the challenges is how to relate the accuracy loss of the estimation with the search effort in the ANNS problem, which relies on the cost analysis of the graph search algorithm~\cite{steiner}.

\textbf{4. Efficient storage compression with dimensionality-reduction techniques.}
As the dimensionality of the vectors grow larger, the storage cost of vectors becomes very high, 
which renders current database storage and query optimization techniques ineffective for the vector type.
While scalar quantization techniques~\cite{lvq} provide an alternative, dimensionality-reduction  offers an orthogonal approach to address this problem.
In this sense, combining these two types of techniques may offer an efficient database storage compression solution for vector data.


\section*{Acknowledgments}
Supported by EU Horizon projects AI4Europe (101070000), TwinODIS (101160009), ARMADA (101168951), DataGEMS (101188416), RECITALS (101168490).

{
\small
\bibliographystyle{plain}
\bibliography{main.bib}
}

 \end{document}